\documentclass[reqno,11pt]{amsart}
\usepackage{amssymb}
\newtheorem{defin}{Definition}[section]

\newtheorem{theorem}{Theorem}

\newtheorem{lemma}{Lemma}

\theoremstyle{remark}

\renewcommand\d{\mathrm d}

\def \Sphere {\mathbb{S}}

\let\Re\undefined \let\Im\undefined
\DeclareMathOperator{\Re}{Re}
\DeclareMathOperator{\Im}{Im}

\DeclareMathOperator{\supp}{supp}

\makeatletter

\newif\ifper\pertrue
\def\per{.}

\newcounter{aucount}
\def\HarvardComma{,}

\newif\ifedplural

\def\ed#1#2{\ifnum\theaucount=0(\fi{#1 #2}\addtocounter{aucount}{1}}
\def\led#1#2{\ifnum\theaucount=0(\edpluralfalse\else\edpluraltrue\fi{#1
    #2} (\editorname.)):\setcounter{aucount}{0}}
\def\editorname{\ifedplural Eds\else Ed\fi}

\def\et{\ifnum\theaucount=1\else\HarvardComma\fi{} and\ }

\def\bti{\@ifnextchar[\bbti\bbbti}
\def\bbti[#1]#2{\emph{#2}, #1,}
\def\bbbti#1{\emph{#1},}

\def\z{\@ifnextchar[\zz\zzz}
\def\zz[#1]#2#3#4#5{\perfalse{#2} \textbf{#3} (#5), #4 [#1]\per}
\def\zzz#1#2#3#4{{#1} \textbf{#2} (#4), #3\ifper\per\fi\pertrue}

\def\pub{\@ifstar\pubstar\pubnostar}
\def\pubnostar{\@ifnextchar[\@@pubnostar\@pubnostar}
\def\@@pubnostar[#1]#2#3#4{#1, #2, #3, #4\per}
\def\@pubnostar#1#2#3{#1, #2, #3\ifper\per\fi\pertrue}
\def\pubstar[#1]#2#3#4{\perfalse #2, #3, #4 [#1]\per}

\makeatother

\def \Real {{\mathbb R}}
\def \Sphere {\mathbb{S}}
\def \Complex {\mathbb{C}}
\def \Natural {{\mathbb N}}
\def \mco  {\mathcal{O}}
\def \Sphere {{\mathbb S}}

\def \Integers {{\mathbb Z}}

\newcommand{\beq}{\begin{equation}}
\newcommand{\eeq}{\end{equation}}
\newcommand{\ba}{\begin{array}}
\newcommand{\ea}{\end{array}}
\newcommand{\bea}{\begin{eqnarray}}
\newcommand{\eea}{\end{eqnarray}}

\newcommand{\R}{\mathbb{R}}

\newcommand{\Z}{\mathbb{Z}}

\newcommand{\N}{\mathbb{N}}

\newcommand{\C}{\mathbb{C}}

\begin{document}

\title[Resonances for Schr\"odinger operators]{Resonances for Schr\"odinger operators
with compactly supported potentials}

\author[T.\ J.\ Christiansen]{T.\ J.\ Christiansen}
\address{Department of Mathematics ,
University of Missouri,
Columbia, Missouri 65211 USA}
\email{tjc@math.missouri.edu}
\author[P.\ D.\ Hislop]{P.\ D.\ Hislop}
\address{Department of Mathematics,
    University of Kentucky,
    Lexington, Kentucky  40506-0027, USA}
\email{hislop@ms.uky.edu}


\begin{abstract}
We describe the generic behavior of the resonance counting
function for a Schr\"odinger operator with a bounded,
compactly-supported real or complex valued
potential in $d \geq 1$ dimensions.
This note contains a sketch
of the proof of our main results \cite{ch-hi1,ch-hi2}
that generically the order of growth of the resonance counting
function is the maximal value $d$ in the odd
dimensional case, and that it is the maximal value $d$
on each nonphysical sheet of the logarithmic Riemann surface in the even dimensional case.
We include a review of previous results concerning
the resonance counting functions for Schr\"odinger operators
with compactly-supported potentials.
\end{abstract}

\maketitle
\thispagestyle{empty}


\section{Introduction and Survey of Results}\label{sec:intro1}

We are interested in the resonance counting function for Schr\"odinger operators
$H_V = - \Delta + V$, acting on $L^2 ( \R^d)$,
with compactly supported real- or complex-valued potentials $V \in L_0^\infty (\R^d)$.
We give a survey of results and discuss on our recent results on the generic
behavior of the resonance counting function \cite{ch-hi1} and \cite{ch-hi2}.
We refer the reader to \cite{polar,ch-hi1, ch-hi2} for complete proofs of
Theorems \ref{th:main1} and \ref{th:main2}.

Resonances are the poles of the meromorphic continuation of the cut-off resolvent
of $H_V$ as we now explain.
For any $V \in L_0^\infty ( \R^d )$, we denote by $\chi_V \in C_0^\infty ( \R^d )$ any
smooth, compactly supported function
such that $\chi_V V = V$. The cut-off resolvent for the Laplacian $H_0 = - \Delta$
is defined to be $\mathcal{R}_0 ( \lambda )
\equiv \chi_V ( H_0 - \lambda^2 )^{-1} \chi_V$.
This is a holomorphic, bounded operator-valued function
on $L^2 ( \R^d)$ for $\lambda \in \C$ for $d \geq 1$ odd,
and for $\lambda \in \Lambda$, for $d \geq 4$ even. For $d=2$ it is
holomorphic on $\Lambda \backslash \{ 0 \}$, with a logarithmic singularity at $\lambda = 0$.
The unperturbed cut-off resolvent $\mathcal{R}_0 ( \lambda)$ has the representation
\beq\label{resolv1}
\mathcal{R}_0(\lambda) = E_1(\lambda) + ( \lambda^{d-2} \log \lambda) \; E_2(\lambda),
\eeq
where $E_1(\lambda)$ and $E_2(\lambda)$ are entire operator-valued functions
with $E_2 (\lambda) = 0$ for $d \geq 1$ odd, and for $d=2$, the operator
$E_2(\lambda=0)$ is a finite-rank operator
This follows, for example, from the
formula for the Green's function for the Laplacian
\beq\label{resolv2}
R_0 (\lambda) = \frac{i}{4} \left( \frac{ \lambda}{2 \pi |x-y|} \right)^{(d-2)/ 2}
H_{(d-2)/2}^{(1)} ( \lambda |x-y|),
\eeq
where the Hankel function of the first kind is defined by $H_\nu^{(1)} (z) = J_\nu (z)
+ i N_\nu (z)$, and the expansion of the Neumann Bessel function $N_\nu (z)$, when $\nu \in \N$.

An application of the second resolvent formula
allows us to write the perturbed, cut-off resolvent $\mathcal{R}_V(\lambda)$ for $H_V$ as
\beq\label{eq:cutoffresolv1}
\mathcal{R}_V(\lambda )( 1 + V \mathcal{R}_0 ( \lambda)) = \mathcal{R}_0 (\lambda) .
\eeq
The resolvent $R_V(\lambda)$ is analytic for $\Im \lambda >> 0$, and it is meromorphic
for $\Im \lambda > 0$ with at most finitely-many
poles with finite multiplicities corresponding to the eigenvalues of $H_V$.
Thanks to (\ref{eq:cutoffresolv1}), the perturbed, cut-off resolvent
$\mathcal{R}_V(\lambda)$ extends as
a meromorphic, bounded operator-valued function on $\C$ if $d \geq 1$ is odd, and onto the
Riemannn surface $\Lambda$ if $\d \geq 4$ is even, with an additional logarithmic
singularity at $\lambda = 0$ when $d = 2$.
The poles of these continuations are the resonances of $H_V$. They are independent of the choice of
$\chi_V$ satisfying the above conditions. It is seen that they correspond
to the values of $\lambda$ for which $1 + V \mathcal{R}_0 (\lambda)$ is not boundedly invertible.
We will use this fact in section \ref{subsec:evendim1}.

We will always assume $V \in L_0^\infty (\R^d)$.
The {\it resonance counting function} $n_V(r)$ for $d \geq 1$ and {\it odd} is defined
to be the number of
poles $\lambda_j$ of $\mathcal{R}_V(\lambda)$,
including multiplicities, with $| \lambda_j | \leq r$ and $\Im \lambda_j < 0$. For $d \geq 2$ even,
the resonance counting function $n_{V,m}(r)$ for the $m^{th}$-sheet $\Lambda_m$, $m \neq 0$,
is defined as the number of poles $\lambda_j$ of $\mathcal{R}_V (\lambda)$
with $| \lambda_j | \leq r$ and $m \pi
< \arg \lambda_j < (m+1) \pi$.

The best results on the asymptotic behavior of $n_V(r)$ is due to Zworski \cite{[Zworski1]} for $d=1$.
He proved that
\beq\label{eq:asympt1}
n_V(r) = \frac{2}{\pi} | \supp V| r + {{o}}(r).
\eeq
He obtained a similar asymptotic expansion in $\R^d$, $d \geq 3$ odd,
for a family of spherically symmetric potentials $V(r)$ with compact support
$[0, a]$ and a discontinuity at $r = a$ \cite{[Zworski2]}. The exact form of the constant
of the leading term in this expansion was identified by Stefanov \cite{stef06}. An asymptotic
expansion of the form (\ref{eq:asympt1}) for a class of super-exponentially decaying potentials
in one-dimension was proved by Froese \cite{froese1} and Simon \cite{simon1}.
In what follows, we consider dimensions $d \geq 2$.

We mention that many results are known for the resonance counting function for Schr\"odinger
operators $H_V (h) = - h^2 \Delta + V$ in the semiclassical regime of small $h > 0$, but
we will not discuss these here. We refer to the extensive papers of Sj\"ostrand \cite{sj1}
and of Sj\"ostrand and Zworski \cite{sj-zw1} for the details and references.
We also refer the reader to two review articles \cite{[Zworski4],[Vodev4]}.
Many aspects of resonances for Schr\"odinger operators were explored by physicists.
Regge \cite{regge1} studied the Jost function and proved that there are infinitely many resonances
for compactly supported, real-valued potentials. This and other aspects of resonances
are discussed by Newton \cite{[newton1]}.
Nussenzveig \cite{[Nussenzveig1]} studied analytically and numerically the poles of the $S$-matrix
for square well and square barrier potentials of the type described in section \ref{sec:example1}.


\subsection{Upper Bounds}\label{subsec:ub1}

In analogy with the celebrated Weyl upper bound for the eigenvalue counting function
for elliptic operators on compact manifolds, upper bounds for the resonance counting
function were proven in the eighties and nineties. Because the meromorphic continuation
of the cut-off resolvent depends on the parity of the dimension, the results
are different for odd and even dimensions.
The upper bounds hold for both complex and real-valued potentials.


\subsubsection{Upper bounds - odd dimensions $d \geq 3$}\label{subsubsec:oddub1}

Polynomial upper bounds on the resonance counting function $n_V(r)$ were first
proved by Melrose \cite{[Melrose1]}.
He showed that $n_V (r) \leq C_d \langle r \rangle^{d+1}$. This upper bound
was improved by Zworski \cite{[Zworski0]} who obtained the optimal
upper bound $n_V (r) \leq C_d \langle r \rangle^d$.
Another proof was given by Vodev \cite{[Vodev1]}
and by Sj\"ostrand and Zworski \cite{sj-zw1}. The optimal constant $C_d$
was computed by Stefanov \cite{stef06}.


\subsubsection{Upper bounds - even dimensions $d \geq 2$}\label{subsubsec:evenub1}

The first upper bounds were proven by Intissar \cite{[Intissar1]}.
Intissar defined, for any $\epsilon > 0$ and $r > 1$,
a resonance counting function $N_{In}(\epsilon, r) \equiv \{ \lambda_j ~|~ r^{-\epsilon} <
| \lambda_j | < r^\epsilon , ~| \arg \lambda_j| < \epsilon \log r \}$.
For even dimensions $d \geq 4$, and for any $\epsilon \in (0 , \sqrt{2} /2 )$, he proved the
polynomial upper bound
$N_{In}(\epsilon, r) \leq C_\epsilon \langle r \rangle^{d+1}$.
Vodev considered the resonance counting function $N_{Vo} (r, a)$ defined as the number of resonances
$\lambda_j$ satisfying $0 < |\lambda_j| < r$ and $| \arg \lambda_j| < a$, for
$r, a > 1$. This function counts the number of resonances in a fixed sector of opening angle
$a$. Vodev proved the upper bound for even dimensions $d \geq 2$ \cite{[Vodev2],[Vodev3]}:
\beq\label{eq:evenub1}
N_{Vo} (r,a) \leq C_d a ( r^d + (\log a)^d ) .
\eeq
Note the explicit dependance of the coefficient on $a$. Since the Green's function has a logarithmic
singularity at zero energy in the two-dimensional case, this required a separate argument.
However, this upper bound does not distinguish between the resonances that might occur on
each sheet $\Lambda_m, m\in \Z^* \equiv \Z \backslash \{ 0 \}$.

\subsection{Lower bounds}\label{subsec:lb1}

There are few lower bounds in dimensions $d \geq 2$. As above, there are
different bounds depending upon
whether the dimension is even or odd. Because of the existence of complex
potentials with no resonances \cite{sownr}, these results
hold only for {\it nontrivial real-valued potentials}.

\subsubsection{Lower bounds - odd dimensions $d \geq 3$}\label{subsubsec:oddLB1}

The best known lower bounds for odd dimensions,
other than the asymptotic results for certain radial potentials
\cite{[Zworski2]},
hold for potentials $V \in L_0^\infty ( \R^d)$ of {\it fixed sign}.
In the odd dimensional case, Lax and Phillips \cite{[LP1]} proved that if $V \geq 0$ is bounded below by
a characteristic function on a ball of radius $R>0$, then
that the number of purely imaginary poles in the lower half complex plane
of modulus less or equal to $r > 0$ is bounded from below by $c_d r^{d-1}$, with $c_d > 0$.
Menzala and Schonbek \cite{menzala-schonbek} allow potentials with a small negative component
so that $- \Delta + V > 0$ in the quadratic form sense.
Vasy \cite{vasy1} proved the existence of an infinite number of purely imaginary poles
when $\pm V \geq 0$, extending \cite{[LP1]} to the case of a strictly negative potential.
These results imply that $n_V(r)$ is similarly bounded from below.
The idea of these proofs
is to note that the scattering amplitude is a monotone function of $V$ at purely imaginary energies.
This reduces the problem to an explicit calculation of the number of purely imaginary
poles for the symmetric step potential $V (x) = V_0 \chi_{B_R (0)} (|x|)$,
for any $V_0 > 0$, where $B_R (x)$ is the ball of radius $R>0$ centered at $x$.

In the general case of nontrivial, smooth, real-valued $V \in C_0^\infty (\R^d)$, Melrose
\cite{lrb} observed from the Poisson formula that
there must be infinitely-many resonances if any of the coefficients in the expansion of the
heat kernel for $k > d-2$ are nonvanishing.
In particular, for $d=3$, there are an infinite number of resonances
since $a_1 = \int V^2 \neq 2$. This was extended to a class of super-exponentially
decaying, smooth, real-valued
potentials in \cite{BanSaB1,SaBZ1,SaBZ2}.
The first quantitative lower bounds for the resonance counting
function for nontrivial, smooth, real-valued $V \in C_0^\infty (\R^d)$, not of fixed sign,
were proved in \cite{TJC1}.
In particular, it was proved there that
\beq\label{eq:lb1}
\limsup_{r \rightarrow \infty} \frac{n_V (r)}{r (\log r)^{-p}} = \infty,
\eeq
for all $p > 1$. For the same family of potentials,
S\'a Barreto \cite{[SaB1]} improved this to
\beq\label{eq:lb2}
\limsup_{r \rightarrow \infty} \frac{n_V (r)}{r} > 0 .
\eeq
We mention that, in particular, all these lower bounds require the potential to be smooth.

\subsubsection{Lower bounds - even dimensions $d \geq 2$}\label{subsubsec:evenLB1}

There are only two results on lower bounds in the even dimensional case for $d \geq 4$.
S\'a Barreto and Tang \cite{SaBT1}
proved the existence of at least one resonance for a real-valued, compactly-supported, smooth nontrivial
potential. S\'a Barreto \cite{[SaB2]} studied the resonance counting function
$N_{SaB} (r)$ defined to be the number of resonances $\lambda_j$
with $1/r < | \lambda_j | < r$ and $| \arg \lambda_j | < \log r$. As $r \rightarrow \infty$,
this region in the Riemann surface $\Lambda$ opens like $\log r$. S\'a Barreto
proved that for even $d \geq 4$,
\beq\label{eq:lbeven1}
\limsup_{r \rightarrow \infty} \frac{N_{SaB}(r)}{ (\log r) ( \log \log r)^{-p}} = \infty,
\eeq
for all $p > 1$.


\subsection{Generic behavior}\label{subsec:generic1}

The {\it order of growth} of a positive, real-valued, monotone increasing function $n(r)$ is defined by
\beq\label{defn:order1}
\rho \equiv \limsup_{r \rightarrow \infty} \frac{ \log n(r) }{ \log r} ,
\eeq
when it is finite. Roughly speaking, we prove that the order of growth of the resonance counting function
is maximal for most real-valued or
complex-valued potentials in $L_0^\infty ( \R^d)$ in the even and odd dimensional cases.
Here, the term ``most potentials'' is meant in the Borel sense. Let $X$ be a metric space. A subset
is said to be {\it Baire typical} or {\it generic} if it is a dense $G_\delta$ subset of $X$.

We refer to the beginning of section \ref{sec:intro1}
for the definition of the resonance counting function $n_V(r)$ in odd dimensions,
and of $n_{V,m}(r)$ for even dimensions.
The {\it order of growth} of the resonance counting function $n_V(r)$ or $n_{V,m}(r)$
is defined by
\beq\label{defn:orderodd1}
\rho_V \equiv \limsup_{r \rightarrow \infty} \frac{ \log n_{V}(r)}{ \log r} , ~~ d ~~\mbox{odd},
\eeq
or, by
\beq\label{defn:ordereven1}
\rho_{V,m} \equiv \limsup_{r \rightarrow \infty} \frac{ \log n_{V,m}(r)}{ \log r}, ~~ d ~~\mbox{even} .
\eeq
As discussed above, it is known that both $\rho_V$ and $\rho_{V,m}$ are bounded above by $d$.
We are interested in the values of these order of growth exponents for {\it generic} potentials.
The notion of generic used here was employed by Simon \cite{simon1}
in his study of singular continuous spectrum for Schr\"odinger operators.
We proved the following theorem in \cite{ch-hi1,ch-hi2}.

\begin{theorem}\label{th:main1}
Let $K \subset \R^d$ be a fixed, compact set with nonempty interior.
There is a dense $G_\delta$ set $\mathcal{V}_F (K) \subset L_0^\infty (K; F)$, for $F=\R$ or $F=\C$, such
that if $V \in \mathcal{V}_F (K)$, and if $d \geq 2$ is even,
then $\rho_{V,m} = d$ for all $m \in \Z \backslash \{ 0 \}$,
or, if $d \geq 3$ is odd, then $\rho_V = d$.
\end{theorem}

For odd dimensions, a similar result holds with $C_0^\infty (K;F)$ in place of
$L_0^\infty (K; F)$ \cite{polar,ch-hi1}.
For the proof of Theorem \ref{th:main1}, it is essential to have
an explicit example of a potential for which the order of growth of the resonance
counting function is $d$. For odd dimensions, Zworski \cite{[Zworski2]} proved the asymptotic
form of $n_V(r)$ for a class of positive, nontrivial, spherically symmetric potentials. This result
is, of course, stronger than the lower bound stated in Theorem \ref{th:main2}.
Since, for our purposes, only a lower bound is required, this gives an alternate, and perhaps simpler, proof.
For the even dimensional case, we proved a lower bound on the resonance counting
function for spherically symmetric, constant, positive potentials on each nonphysical sheet.
Let $B_R(x)$ be the ball of radius $R>0$ centered at $x \in \R^d$.

\begin{theorem}\label{th:main2}
Let $V(x) = V_0 \chi_{B_R(0)} ( x)$ for $V_0 > 0$.
Then, for $d$ even, there is a constant $c_m > 0$ so that
$n_{V , m}(r) \geq c_m r^m$, for $m \in \Z^*$. For $d$ odd, there is a constant $c_0 > 0$
so that $n_V(r) > c_0 r^d$.
\end{theorem}

In the remainder of this article, we sketch the main ideas behind the proofs of
Theorems \ref{th:main1} and \ref{th:main2}. Section \ref{sec:reduction1} presents the
reduction to a problem of counting zeros of a function holomorphic on a half-plane.
We prove Theorem \ref{th:main1} in section \ref{sec:genericbehavior1}
modulo results concerning plurisubharmonic functions that are presented
in sections \ref{sec:pluri1} and \ref{sec:parampot1}. The proof of Theorem \ref{th:main2}
is sketched in section \ref{sec:example1}.

\section{Reduction to a zero counting problem}\label{sec:reduction1}

As is typical for these problems, we reduce the estimate on the resonance counting
function to one on the number of zeros of a function holomorphic in the half plane $\Im \lambda
> 0$. In our papers, we took different approaches depending upon whether the dimension is even or odd.
Here, we unify the approaches and show how to use the method of \cite{ch-hi2} for either parity of the
dimension.
For odd dimensions, in \cite{ch-hi1} and \cite{polar},
resonances were treated by considering the $S$-matrix and its scattering phase.
Crucial for the argument is the fact that the $S$-matrix admits a
Weierstrass product representation
in the complex plane. For the even dimensional case, we work directly with the
second resolvent formula and we do that here for any dimension.


\subsection{Resolvent approach: any dimension}\label{subsec:evendim1}

Let $V \in L_0^\infty ( \R^d)$ and let $\chi_V$
be any compactly-supported, smooth function $\chi_V \in C_0^{\infty}(\R^d)$
so that $\chi_V V = V$.
The cut-off free resolvent $\mathcal{R}_0 ( \lambda) \equiv \chi_V R_0(\lambda)\chi_V $ has a
meromorphic continuation to $\Lambda$,
the logarithmic cover of the plane, for $d \geq 4$ even, or to $\C$,
for $d \geq 1$, odd (see, for example, \cite{{lrb},[ST1]}). For $d=2$
there is a logarithmic singularity at $\lambda =0$. For $d$ odd, there are two distinct sheets,
the physical sheet $\Lambda_0$, and $\Lambda_{-1}$. For $d$ even, there is the physical sheet
and infinitely-many distinct nonphysical sheets $\Lambda_m$, $m \in \Z^*$.
In order to unify notation, we define $m(d)$
to be equal to $m \in \Z$ when $d$ is even, and, when $d$ is odd, we take $m = -1$
to be the lower-half complex plane, so $m(d)$ is $m \mbox{mod} ~2$. In this way,
the sheets $\Lambda_{2k}$ are all identified with $\Lambda_0$ and the sheets $\Lambda_{2k+1}$ are all identified with
$\Lambda_{-1}$.

We use the following
key identity, that follows from (\ref{resolv2}) and the formulas for the meromorphic continuation
of Hankel functions (see \cite[section 6]{ch-hi2} or \cite[chapter 7]{[Olver1]}), relating the free resolvent
on $\Lambda_m$ to that on $\Lambda_0$, for any $m \in \Z$,
\begin{equation}\label{eq:difference}
R_0(e^{im\pi}\lambda)=R_0(\lambda)-m(d) T(\lambda), ~~\mbox{where} ~m(d) = \left\{ \begin{array}{ll}
  m \mod 2 & d ~\mbox{odd} \\
  m & d ~\mbox{even} .
  \end{array}
   \right.
\end{equation}
The operator $T(\lambda)$ on $L^2 (\R^d)$ has a Schwartz kernel
\begin{equation}\label{Top1}
T(\lambda, x,y)= i \pi (2 \pi)^{-d} \lambda^{d-2}
\int_{\Sphere^{d-1} }e^{i\lambda (x-y)\cdot \omega}d\omega ,
\end{equation}
see \cite[Section 1.6]{lrb}.
We note that for any $\chi \in C_0^{\infty}(\R^d)$, $\chi T(\lambda)
\chi$ is a holomorphic trace-class operator for $\lambda \in \C$.
The operator $T$ has a kernel proportional to $|x-y|^{(-d+2)/2} J_{(d-2)/2}(\lambda |x-y|)$ when $d$ is odd,
and to $|x-y|^{(-d+2)/2} N_{(d-2)/2}(\lambda |x-y|)$ when $d$ is even. The different behavior
of the free resolvent for $d$ odd or even is encoded in (\ref{eq:difference}).

By the second resolvent formula (\ref{eq:cutoffresolv1}),
the poles of $\mathcal{R}_V (\lambda)$
with multiplicity, correspond to the
zeros of $I+VR_0(\lambda)\chi_V$.
We can reduce the analysis of the zeros
of the continuation of $I+VR_0(\lambda)\chi_V$ to $\Lambda_m$
to the analysis of zeros of a related operator on $\Lambda_0$ using (\ref{eq:difference}).
If $0<\arg \lambda<\pi$ and $m\in \Z$, then
$e^{im\pi} \lambda \in \Lambda_m$, and
\begin{align*}
I+VR_0(e^{im \pi}\lambda)\chi & = I+V(R_0(\lambda)-m T(\lambda))\chi_V \\
& = (I+VR_0(\lambda)\chi_V)(I-m (I+VR_0(\lambda)\chi_V)^{-1}V T(\lambda)\chi_V).
\end{align*}
For any fixed $V\in L^{\infty}_0 (\R^d)$,
there are only finitely many poles of $(I+VR_0(\lambda) \chi_V)^{-1}$
with $0<\arg \lambda<\pi$.  Thus
\begin{equation}\label{eq:fmfirst}
f_{V,m} (\lambda)=\det(I -m (I+VR_0(\lambda)\chi_V)^{-1}V T(\lambda)\chi_V)
\end{equation} is
a holomorphic function of $\lambda$
when $0<\arg \lambda<\pi$ and
$|\lambda|>c_0 \langle \|V\|_{L^{\infty}}\rangle$.
Moreover, with at most a finite number of exceptions, the zeros
of $f_{V,m}(\lambda)$, with $0<\arg \lambda<\pi$ correspond, with multiplicity,
to the poles of $\mathcal{R}_V (\lambda)$ with $m \pi <\arg \lambda < (m+1)\pi$.
Henceforth, we will consider the function $f_{V,m} (\lambda)$,
for $m \in  \Z^* \equiv \Z \backslash \{ 0 \}$, on $\Lambda_0$.
For $d$ odd, we are only interested in $m = -1$. In this case, the zeros
of $f_{V , -1} (\lambda)$, for $\lambda \in \Lambda_0$, correspond to the resonances.
A similar approach in the $d$ odd case was employed by Froese \cite{froese1}.


\subsection{The $S$-matrix approach in odd dimensions}\label{subsec:odddim1}

We comment on the case of odd dimensions
used in \cite{polar, ch-hi1}.
We denote the scattering matrix for the pair $H_0 = - \Delta$
and $H_V = H_0+V$ by $S_V(\lambda)$, acting on
$L^2 ( \Sphere^{d-1})$. In the case
that $V$ is real-valued, this is a unitary operator for $\lambda \in \R$.
The $S$-matrix is given explicitly by
\begin{equation}
\label{eq:sm}
S_V(\lambda)=I+c_d \lambda^{d-2} \pi_{\lambda}
(V-VR_V(\lambda)V)\pi^t_{-\lambda} \equiv I + \mathcal{T}_\lambda,
\end{equation}
where
$\pi_\lambda : L^2 (\R^d) \rightarrow L^2 ( \Sphere^{d-1})$ is defined by $(\pi_{\lambda}f)
(\omega)=\int e^{-i\lambda x\cdot \omega}f(x) ~dx$ \cite{[Yafaev]}.
Under the assumption that $\mbox{supp}~V$ is compact, the
operator $\mathcal{T}_\lambda : L^2 ( \Sphere^{d-1}) \rightarrow L^2 ( \Sphere^{d-1} )$ is
trace class.
The $S$-matrix has a meromorphic continuation to the entire
complex plane with finitely many poles for $\Im \lambda > 0$ corresponding to eigenvalues
of $H_V$, and resonance poles in $\Im \lambda < 0$.
We recall that if $\Im \lambda_0 \geq c_0 \langle \|V\|_{L^{\infty}} \rangle$,
the multiplicities of $\lambda_0$, as a zero of $\det S_V(\lambda)$, and
of $-\lambda_0$, as a pole of the cut-off resolvent $\mathcal{R}_V (\lambda)$, coincide.
Consequently, the function
\beq\label{eq:oddfnc1}
f_V(\lambda) \equiv \det S_V(\lambda),
\eeq
is holomorphic for $\Im  \lambda >
c_0 \langle \| V \|_{L^\infty} \rangle$,
and well-defined for $\Im \lambda \geq 0$ with finitely many poles.
Hence, the problem of estimating the number of zeros of $f_V(\lambda)$ in the upper
half plane is the same as estimating the number
of resonances in the lower half plane.

This is facilitated in the odd dimensional case by the well-known representation
of $f_V(\lambda)$ in terms of canonical products. Let $G(\lambda; p)$ be defined
for integer $p \geq 1$, by
\beq\label{eq:product1}
G( \lambda ; p ) = (1-\lambda)e^{\lambda + \lambda^2/ 2 + \cdots + \lambda^p/p},
\eeq
and define
\beq\label{eq:poly1}
P(\lambda) = \Pi_{\lambda_j \in \mathcal{R}_V, \lambda_j \neq 0} G(\lambda / \lambda_j ; d-1).
\eeq
Then the function $f_V(\lambda)$ may be written as
\beq\label{eq:defnfvodd1}
f_V(\lambda) = \alpha e^{i g(\lambda) } \frac{P(-\lambda)}{P(\lambda)} ,
\eeq
where $g(\lambda)$ is a polynomial of order at most $d$.
Careful study of the scattering matrix and the upper bound of section \ref{subsubsec:oddub1}
may be used to show that $f_V(\lambda)$ is of order at most $d$ in the half-plane $\Im \lambda > c_0 \langle
\| V \|_\infty \rangle$, see \cite{[Zworski3]}.
It is the representation (\ref{eq:defnfvodd1}) that is not available in the even
dimensional case.
This facilitates the construction of a plurisubharmonic function
(see sections \ref{sec:pluri1}--\ref{sec:parampot1})
in the odd dimensional case as done in \cite{polar,ch-hi1}.


\section{Proof of Generic Behavior}\label{sec:genericbehavior1}

The proof of Theorem \ref{th:main1} consists of two components. The first is
the $G_\delta$-property of the set of all potentials in $L_0^\infty ( K; F)$, for
$F = \R$ or $F = \C$ separately, and $K \subset \R^d$ a compact set with nonempty interior,
with the correct order of growth. This is fairly easy to prove. The second is the proof that
this $G_\delta$-set is dense.
The density argument
is the difficult part.

Many of the statements in the proof of the $G_\delta$-property do
not depend on whether $d$ is even or odd.
We write $f_{V, m}(\lambda)$ with the understanding that $m \in \Z^*$ for $d$ even,
and $m = -1$ for $d$ odd. We use the same convention for the counting function
$n_{V,m}(r)$, so that $n_{V, -1}(r) = n_V(r)$.

We first need a Jensen-type theorem that relates the order of growth of the zero counting function
$n_{V,m}(r)$ to the asymptotic behavior of $f_{V,m} (\lambda)$. In general, we consider a function $h$
holomorphic in
the set $\{ \lambda \in\Complex:\; |\lambda|\geq R \geq 0,\; \Im \lambda \geq 0\}$.
For $r>R$, we
define $n_{+,R}(h,r)$ to be the number of zeros of $h$, counted with multiplicities, in the closed
upper half plane with norm between $R$ and $r$, inclusive.

\begin{lemma}\label{l:horder}
Let $R>0$, let $h$ be holomorphic in
 $\{ \lambda \in \C ~|~ |\lambda| \geq R > 0, \Im \lambda \geq 0 \}$.
Suppose, in addition, that $h$ has only finitely many
zeros on the real axis.
Suppose that for some $p>0$, and for some $\epsilon>0$, we have
$$
\int_R^{r}\left|\frac{h'(s)}{h(s)}\right| ds
= \mathcal{O}(r^{p-\epsilon})\; \text{and}\;
\int_{-r}^{-R}\left|\frac{h'(s)}{h(s)}\right| ds
= \mathcal{O}(r^{p-\epsilon}).
$$
Then $n_{+,R}(h,r)$ has order $p$ if and only if
$$\lim \sup _{r\rightarrow \infty}
\frac{\log \int_0^{\pi}\log|h(re^{i\theta})|d\theta}{\log r}=p.$$
\end{lemma}

We will apply this lemma with $h$ taken to be $f_{V,m}$.
For positive constants $N,M,q > 0$ and $j > 2 N$, we define subsets of
$L^\infty_0 ( K; F)$ by
\bea\label{sets1}
A_m (N,M,q,j) & \equiv & \left\{ V \in L^\infty_0 (K; F) ~:~ \langle \| V \|_{L^\infty}
\rangle \leq N, \right. \nonumber \\
 & & \int_0^\pi ~\log | f_{V,m} ( r e^{i \theta} ) | ~d \theta
\leq M  r ^q, \nonumber \\
 & & \left. \mbox{for}\;  2N  \leq r \leq j \right\} .
\eea

\begin{lemma}\label{closed1}
The set $A_m (N,M,q,j) \subset L_0^\infty (K;F)$ is closed.
\end{lemma}

The proof is a continuity argument in the potential $V$. Suppose
$V_k \in A_m (N,M,q,j)$ converges to $V$ in the $L^\infty$-norm. We show that
the corresponding functions $f_m (z)$ converge using the basic bound \cite{[Simon2]}
\beq\label{eq:trace1}
| \det ( 1 + A) - \det (1 + B) | \leq \| A - B\|_1 e^{ \|A\|_1 + \|B\|_1 + 1} ,
\eeq
with $A = (I + V_j R_0 (\lambda) \chi)^{-1} V_j T(\lambda) \chi$ and
$B= (I + V R_0 (\lambda) \chi)^{-1} V T(\lambda) \chi$.

In the next step, we characterize
those $V \in L^\infty_0 (K; F)$ for which the resonance counting function exponent
is strictly less than the dimension $d$.
For $N,\;M,\;q>0$, let
\beq
B_m (N,M,q)= \bigcap_{j\geq 2N }A_m (N,M,q,j).
\eeq
Note that $B_m (N,M,q)$ is closed by Lemma \ref{closed1}.

\begin{lemma}\label{l:inB}
Let $V\in L^{\infty}_0 (K;F)$, with
\beq\label{eq:inB1}
\limsup _{r\rightarrow
\infty}\frac{\log n_{V,m} (r)}{\log r}<d.
\eeq
Then there exist $N,\; M\in \Natural$,
$l\in \Natural$, such that $V\in B_m (N,M,d-1/l)$.
\end{lemma}

The proof of this lemma uses Lemma \ref{l:horder} with $h = f_{V,m}$. Condition
(\ref{eq:inB1}) implies that
\beq\label{eq:inB2}
\limsup_{r\rightarrow \infty}\frac{\log \int_0^\pi
\log| f_{V,m} (   re^{i\theta})|
~d \theta }{\log r}
=p<d.
\eeq
From this it follows that there are constants $p', M >0$, with $p \leq p' <d$,
so that
$$
\int_0^\pi \log | f_{V,m} (r e^{i \theta})| ~d \theta \leq M r^{p'}
$$
when $r \geq c_0 \langle \|V\|_{L^\infty} \rangle$. The lemma then follows by choosing $\ell \in \N$
so that $p' \leq d - 1 / \ell$.

\begin{lemma}
\label{l:gd}
For $d \geq 2$ even and $m \in \Z^*$ or $d \geq 1$ odd and $m = -1$, the sets
$$\mathcal{M}_m = \left\{ V\in L^{\infty}_0(K;F): \;
\lim \sup_{r\rightarrow \infty}\frac{\log n_{V,m}(r)}{\log r}=d \right\}, $$
are $G_\delta$-sets.
Furthermore, for $d \geq 2$ even, the set
$$\mathcal{M}=\bigcap_{m\in \Z^*}\mathcal{M}_m$$
is also a $G_\delta$-set.
\end{lemma}
\begin{proof} Lemma \ref{l:inB} shows that the complement of
$\mathcal{M}_m$ is contained in
\begin{equation*}
\bigcup_{(N,M,l) \in \Natural^3}  B_m (N,M,d-1/l),
\end{equation*}
which is an $F_{\sigma}$ set since it is a countable union of
closed sets.  By Lemma \ref{l:horder}, if $V\in \mathcal{M}_m,$
then $V\not \in B_m(N,M,d-1/l)$ for any $N,\; M,\; l \in \Natural.$
Thus $\mathcal{M}_m$ is the complement of an $F_{\sigma}$ set.
\end{proof}

\noindent
Theorem \ref{th:main1} follows from Lemmas \ref{l:horder} - \ref{l:gd} and the density argument
presented here that relies on sections \ref{sec:pluri1}--\ref{sec:example1}.

\begin{proof}[Proof of Theorem \ref{th:main1}]
Since Lemma \ref{l:gd} shows that $\mathcal{M}_m$
is a $G_{\delta}$ set,
we need only show that each is
dense in $L^{\infty}_0 (K;F)$. To do this,
we follow the proof of \cite[Corollary 1.3]{polar} with appropriate modifications.
Let $V_0\in L^{\infty}_0 (K;F)$ and let $\epsilon >0$.
By Theorem \ref{th:main2}, proved in section \ref{sec:example1},
we may choose a nonzero, real-valued, spherically symmetric $V_1\in
L^{\infty}_0 (K;\Real)$ so that $V_1\in \mathcal{M}_m$, for $m \in \Z^*$.
We consider the holomorphic function $V(z)=V(z,x)=zV_1(x)+(1-z)V_0(x)$, for $z \in \C$.
This function satisfies Assumptions (V) of section \ref{sec:parampot1},
with $V(1) = V_1$ and $V(0)=V_0$.
Thus, by \cite[Theorem 1.1]{polar}, for $d$ odd, and by \cite[Theorem 3.8]{ch-hi2}, for $d$ even,
there exists a pluripolar set $E_m \subset \Complex$ (see Definition \ref{pluripolar1}),
so that for $z\in \Complex\setminus E_m$, we have
$$
\limsup_{r\rightarrow \infty}\frac{\log n_{V(z),m}(r)}{\log r}=d.
$$
If we set $E = \cup_{m \in \Z^*} E_m$,
the set $E \subset \C$ is also pluripolar \cite[Proposition 1.37]{l-g}.
Since ${E} \cap \R \subset \R$ has
Lebesgue measure $0$
(e.g. \cite[Section 12.2]{ransford}),
we may find $z_0\in \Real$, $z_0\not \in E$, with
$|z_0|<\epsilon / (1+ \|V_0\|_{L^{\infty}}+\|V_1\|_{L^{\infty}})$.
Then $V(z_0)\in \mathcal{M}_m$
for all $m \in \Z^*$,
and $\|V(z_0)-V_0\|_{L^{\infty}}<\epsilon$.  Moreover, if $V_0$ is real-valued,
so is $V(z_0)$.
\end{proof}

\section{Plurisubharmonic functions and pluripolar sets}\label{sec:pluri1}

A key role is played in the density part of the
proof of Theorem \ref{th:main1} by the theory of plurisubharmonic functions. Plurisubharmonic functions
were first used in the study of the resonance counting function
in \cite{polar}.
A basic reference is the book by Lelong and Gruman \cite{l-g}.
We refer to an open connected set $\Omega \subset \C^{k}$ as a domain.
For two domains $\Omega'$ and $\Omega$, we use the notation $\Omega' \Subset \Omega$
if $\overline{\Omega'}$ is compact and $\overline{\Omega'} \subset \Omega$.

\begin{defin}\label{defn:plurisubharm1}
A real-valued function $\phi (z)$ taking values in $[ -\infty, \infty)$ is
{\em plurisubharmonic} in
a domain $\Omega \subset \C^k$, and we write $\phi \in PSH ( \Omega)$, if:
\begin{itemize}
\item $\phi$ is upper semicontinuous and $\phi \not\equiv - \infty$;
\item for every $z \in \Omega$, for every $w \in \C^k$,  and every $r > 0$ such that
$\{ z + u w : |u| \leq r, u \in \C \}
\subset \Omega$
we have
\beq\label{subMV1}
\phi (z) \leq \frac{1}{2 \pi} \int_0^{2 \pi} ~\phi ( z + r e^{i \theta}w ) ~d \theta .
\eeq
\end{itemize}
\end{defin}

If the dimension $k=1$, then this is just the definition of a subharmonic function.
A basic example of a PSH function is $\phi_f (z) = \log | f(z)|$ where $f$ is holomorphic in a domain
$\Omega$. If $f(z_0) = 0$, for some $z_0 \in \Omega$, then $\phi (z_0) = - \infty$. This shows that there
is a connection between the behavior of this PSH function $\phi_f (z)$ on $\Omega$ and
the zeros of $f$ on $\Omega$. The points where a nontrivial PSH function takes the value $- \infty$ are special and rare.

\begin{defin}\label{pluripolar1}
A set $E \subset \C^k$ is {\em pluripolar} if for each $a \in E$ there is a neighborhood
$V_a$ containing $a$ and a function $\phi_a \in PSH(V_a)$ such that $E \cap V_a \subset \{
z \in V_a : \phi_a (z) = - \infty \}$.
\end{defin}

Pluripolar sets have many properties. Of course, if $E \subset \C$ is pluripolar,
then the two-dimensional Lebesgue measure of $ E $ is zero.
Moreover, the one-dimensional Lebesgue measure
of $E \cap \R$ is also zero.

The order of growth $\rho$ of a PSH function $\phi$ on the complex plane
$\C$ is defined to be
\beq\label{eq:orderpsh1}
\rho \equiv \limsup_{r \rightarrow \infty} \frac{\log \sup_{0 \leq \theta \leq 2 \pi }
| \phi ( re^{i \theta} )|}{\log r},
\eeq
when it exists and is finite.
We consider PSH functions on domains of the form $\Omega \times \C$, with $\Omega
\subset \C^k$, $k \geq 1$.
For fixed $z' \in \Omega$, let  $\rho (z')$ be the order
of growth of $u \in \C \rightarrow  \phi (z' , u)$,
where $z = (z' , u) \in \Omega \times \C$.

We use two main results in the theory of PSH functions in order to analyze PSH
functions on domains of the form $\Omega \times \C$ and their parameterized order of growth
$\rho(z')$. These are presented in the textbook by Lelong and Gruman \cite{l-g}.
The first result (\cite[Proposition 1.40]{l-g})
is that for $\Omega' \Subset \Omega$, there is a sequence of negative PSH functions $\psi_n $ on $\Omega'$
so that $[\rho (z')]^{-1} = - \limsup_{n \rightarrow \infty} \psi_n (z')$, where $z' \in \Omega'$.
The second result (\cite[Proposition 1.39]{l-g})
concerns a sequence $\psi_n$ of PSH functions uniformly bounded above
on $\Omega \subset \C^k$ with $\limsup_{n \rightarrow \infty}  \psi_n \leq 0$.
If there is one $z_0 \in \Omega$ such that $\limsup_{n \rightarrow \infty} \psi_n (z_0) =0$,
then the set $\{ z \in \Omega ~|~ \limsup_{n \rightarrow \infty} \psi_n (z) < 0 \}$ is pluripolar
in $\Omega$.

In our application of these results, we will show that the order of growth of
a certain PSH function is at most $d$ so that, applying the first result sketched above,
$\limsup_{n \rightarrow \infty} ( \psi_n (z') + 1/d ) \leq 0$.
We will find a $z_0$ for which the limit superior is exactly zero, so that the second result
mentioned above implies that the limit superior is zero for all $z \in \Omega'$ except for a
pluripolar set. This is the essence of the density argument.
We next turn to the construction of the appropriate PSH functions.


\section{Parameterized potentials and the order of growth}\label{sec:parampot1}

The theory of plurisubharmonic functions is applied to the resonance counting problem
through the introduction of families of holomorphic potentials $V(x;z)$ with $z \in \Omega \subset \C^k$.
The main result is roughly the following: If there is a point $z_0 \in \Omega$
for which the order of growth $\rho_{V(z_0)}$ for the
resonance counting function for the potential $V(z_0)$ is maximal (so upper and lower bounds on
the resonance counting function of the same order $d$ are required), then
there is a pluripolar set $E \subset \Omega$ so that the order of growth is maximal
for all potentials $V(z)$ with $z \in \Omega \backslash E$. This provides the proof of density
as in section \ref{sec:genericbehavior1}.

In order to achieve this result, one has to construct a plurisubharmonic function (PSH)
that reflects the order of growth of the resonance counting function for the family of potentials
$V(z;x)$. Let $\Omega \subset \C^k$ be an open connected set.
Let $V(z) \equiv V(z;x)$ be a family of potentials satisfying
\begin{itemize}
\item For $z\in \Omega$, $V(z,\cdot )\in L^{\infty}_0 (\Real^d)$.
\item  The function $V(z,x)$ is holomorphic in $z\in \Omega$.
\item There is a compact set $K_1 \subset \R^d$ so that
for $z\in \Omega$, $V(z,x)=0$ if $x\in \Real^d\setminus K_1$.
\end{itemize}
We will refer to these properties as {\it Assumptions (V)}. We are interested in the resonance
counting functions for potentials satisfying Assumptions (V).
For such a potential $V(z)$, we define, in analogy to (\ref{eq:fmfirst}), the function
\begin{equation}\label{eq:fmz}
f_m(z,\lambda)=\det(I -m (I+V(z) R_0(\lambda)\chi)^{-1} V(z) T(\lambda) \chi) ,
\end{equation}
for $\lambda \in \Lambda_0$.

In order to apply the theory of \cite{l-g},
we construct a plurisubharmonic function $M(z, u)$ on $\Omega' \times \C$,
for any $\Omega' \Subset \Omega$,
from
(\ref{eq:fmz}). This function
has the property that $M(z, |u|)$ is a positive, monotone increasing, function of $|u|$,
for any $z \in \Omega$.
Recalling the holomorphicity of $f_m$, we define, for any $\epsilon > 0$, a function $g_{m,\epsilon}(z,u)$
by
\begin{equation}\label{eq:gmedef}
g_{m,\epsilon}(z,u)=\int_0^{\pi}\log |f_m(z,ue^{i\theta})|d\theta
+ \log |e^{u^{d-\epsilon}}|.
\end{equation}
For $\Omega'\Subset \Omega$, we define the constant
$V_{M,\Omega'}=\max _{z\in \Omega'}\|V(z, \cdot)\|_{L^{\infty}}.$
The function $g_{m,\epsilon}(z,u)$ is PSH on the strip-like domain of the form
$\Omega' \times U_{\Omega'}$, where $\Omega' \Subset \Omega$ and $U_{\Omega'}
\subset \C$
is given by
\beq\label{eq:strip1}
U_{\Omega'} = \{ u \in \C ~|~ |\Im u | < 2 , \Re u > c_0 V_{M, \Omega'} , | \arg u | < \pi /4 \}.
\eeq
The second term in (\ref{eq:gmedef}) allows us to control the location of the maximum of $g_{m,\epsilon}$
on the intersection of the domain $\Omega ' \times U_{\Omega'}$ with the arcs $|u| = r$.

Starting with $g_{m, \epsilon}$ defined in (\ref{eq:gmedef}),
we construct a PSH function on $\Omega' \times \C$, for any $\Omega' \Subset \Omega$,
through a series of extensions that we briefly summarize here.
First, one proves that there exists $r_m > 0$ so that for $r >
r_m \langle V_{M,\Omega'} \rangle^{1/(1-\epsilon)}  > 0$,
\beq\label{eq:gmax1}
\max_{\substack{ |\Im u| \leq 1, \; \Re u>0 \\
 |u|= r,\; u\in U_{\Omega'}}}
g_{m,\epsilon}(z,u)
>  \max _{\substack{\Im u = \pm 1, \; \Re u>0 \\
 |u|= r,\; u\in U_{\Omega'}}}
g_{m,\epsilon}(z,u),
\eeq
for $z \in \Omega'$.
%
Second, we note that by using inequality (\ref{eq:gmax1}) we have
\beq\label{eq:defnmax1}
\tilde{M}_{m,\epsilon,\Omega'}(z,w)=
\max_{\substack{|\Im u |\leq 1, \; |\arg u|\leq \pi/4\\
 r_m(\langle V_{M,\Omega'}\rangle^{1/(1-\epsilon)}+1)\leq
|u| \leq |w|}}
g_{m,\epsilon}(z,u) ,
\eeq
is plurisubharmonic on $\Omega' \times
\{ w\in \Complex: \;  \tilde{r}_{m,\epsilon}(\Omega',V)<|w|\}$, for a suitable constant
$\tilde{r}_{m,\epsilon}(\Omega',V) >0$.
Finally, we extend this to a PSH function on $\Omega' \times \C$.
We prove that the function
\beq\label{eq:devenpsh2}
M_{m,\epsilon,\Omega'}(z,w)=
\left\{ \begin{array}{l l}
\max (1,\tilde{M}_{m,\epsilon, \Omega'}(
z,\tilde{r}_m(\Omega',V)
+1) , & \text{if}\;
|w|\leq \tilde{r}_m(\Omega',V)
+1\\
\max (1,\tilde{M}_{m,\epsilon,\Omega'}(z,w)),& \text{if}\; |w|\geq
\tilde{r}_m(\Omega',V),
\end{array}
\right.
\eeq
is plurisubharmonic on $\Omega'\times \Complex$.

Note that the
dependence of $M_{m, \epsilon, \Omega'}(z,w)$ on $w$ is only through
the norm $|w|$
and is a monotone increasing function
of $|w|$.
The following lemma
demonstrates the relationship between
the order of $n_{V(z)}(r)$ and the order of $r\mapsto M_{m, \epsilon, \Omega'}(z,r)$
for any $z \in \Omega'$ fixed.
\begin{lemma}\label{l:morder}
Let $\Omega'\Subset\Omega$
and let $\rho_{m, \epsilon, \Omega'}(z)$ be the order
of $r\rightarrow M_{m, \epsilon, \Omega'}(z,r)$.
We then have
\beq\label{eq:order3}
\rho_{m, \epsilon, \Omega'}(z)= \max(d-\epsilon,
\text{order of}\; n_{V(z),m}(r))
\eeq
for $z\in \Omega'$, where, as above, $n_{V(z),m}(r)$ is the number of
resonances of $H_{V(z)}$ on $\Lambda_m$,
$m \in \Z^*$  of norm at most $r>0$.
\end{lemma}

The main result of this construction, and the general results for PSH functions
presented in section \ref{sec:pluri1}, is the following theorem. The theorem assets that
except for a small set of $z \in \Omega$, the order of growth of $n_{V(z),m}(r)$ is $d$
provided it is bounded by $d$ for $z \in \Omega$ and
provided that it obtains this value for at least one point in $\Omega$.

\begin{theorem}\label{main-psh1}
Let $\Omega\subset \C^{d'}$ be an open connected
set, let $m\in \Integers$, and let $V(z,x)$ satisfy the assumptions
(V).  If for some $z_m\in \Omega$, the function $n_{V(z_m),m}(r)$ has order
$d$, then there is a pluripolar set $E_m\subset \Omega$
such that $n_{V(z),m}(r)$ has order $d$ for $z\in \Omega \setminus E_m$.
Moreover, if for each $m\in \Integers^*$, there is a $z_m$ such that
$n_{V(z_m),m}(r)$ has order
$d$, then there is a pluripolar set $E$ such that for every $m\in \Integers^*$,
the function $n_{V(z),m}(r)$ has order $d$ for $z\in \Omega \setminus E$.
\end{theorem}


\section{Lower bounds for some spherically symmetric potentials}\label{sec:example1}

We compute a lower bound on the number of resonances for $H_V$ when $V(x) = V_0 \chi_{B_R (0)} (x)$,
with $V_0 > 0$, using separation of variables and uniform asymptotics of Bessel and Hankel functions
due to Olver \cite{[Olver1],[Olver2],[Olver3]}.
This method works for $d$ even or odd, thus providing an alternate to using
the more precise asymptotic result of Zworski \cite{[Zworski2]} for $d$ odd,
as was done in \cite{polar,ch-hi1}.
Because of the spherical symmetry of $V$, we can reduce
the Hamiltonian $H_V$ to a direct sum of Hamiltonians $H_\ell$, for $\ell = 0 , 1, 2, \ldots$
acting on $L^2 (\R^+)$. An important parameter is $\nu \equiv \ell + (d - 2) / 2$ that
is an integer for $d$ even and half an odd integer for $d \geq 3$ odd.
We construct the Green's function
on the physical sheet $\Lambda_0$ for the reduced Hamiltonian $H_\ell$.
We let $V(r) =
V_0 \chi_{[0,1]} (r)$, with $V_0 > 0$, and we let $\Sigma (\lambda) \equiv ( \lambda^2 -
V_0)^{1/2}$, where the square root is defined so that this function has
branch cuts $( - \infty, -V_0^{1/2}] \cup [ V_0^{1/2}, \infty)$.
Because of the simple nature of the potential $V(r)$,
the reduced ordinary differential equation for $0 < r < 1$ is
\beq\label{reduced1}
- \psi_\nu '' - \frac{(d-1)}{r} \psi_\nu ' + \frac{\ell ( \ell + d -2)}{r^2}
  \psi_\nu
  = \Sigma(\lambda)^2 \psi_\nu ,
\eeq
and for $r >1$, the solution $\psi_\nu$ satisfies the free equation
\beq\label{reduced2}
- \psi_\nu '' - \frac{(d-1)}{r} \psi_\nu ' + \frac{\ell ( \ell
+ d -2)}{r^2} \psi_\nu
  = \lambda^2 \psi_\nu.
\eeq

We use the ordinary Bessel and Hankel functions of the first kind,
denoted by $J_\nu$ and $H_\nu^{(1)}$, and their spherical counterparts $j_\nu$ and $h_\nu^{(1)}$.
We refer to \cite{[Olver1]} for complete definitions and properties.
We choose two linearly independent solutions,
$\phi_{\nu}$ and $\psi_{\nu}$ of (\ref{reduced1})--(\ref{reduced2})
so that
$\phi_\nu ( r=0; \lambda) = 0$ and $\psi_\nu (r; \lambda) = h_\nu ( \lambda r)$
for $r >1$. The Green's function has the form
\beq\label{green1}
G_\nu (r,r'; \lambda) =
\frac{1}{W_\nu(\lambda) } \left\{ \begin{array}{ll}
     \phi_\nu ( r; \lambda  ) \psi_\nu ( r' ; \lambda ) , &  r < r' \\
    \phi_\nu ( r' ; \lambda ) \psi_\nu ( r; \lambda ) , &  r > r'
     \end{array}
     \right. ,
\eeq
where the Wronskian $W_\nu (\lambda)$, evaluated at $r=1$, is given by
\beq\label{wronskian1}
W_\nu (\lambda) = \Sigma (\lambda) j_{\nu} ' ( \Sigma (\lambda) ) h_{\nu }^{(1)} (\lambda ) -
\lambda j_\nu ( \Sigma ( \lambda)) h_{\nu}^{(1) '}  (\lambda ) .
\eeq

It follows that
$\lambda_0 \in \Lambda_m$, $m \neq 0$, for $d$ even,
or $\lambda_0 \in \Lambda_{-1} \equiv \Complex^-$, if $d$ is odd, is a resonance if it satisfies the
following condition:
\beq\label{basic11}
\Sigma (\lambda_0) J_{\nu} ' ( \Sigma (\lambda_0) ) H_{\nu}^{(1)} (\lambda_0 ) -
\lambda_0 J_\nu ( \Sigma (\lambda_0))
H_{\nu}^{(1) '}  (\lambda_0 ) = 0, ~\nu = \ell + (d-2)/2 ,
\eeq
where we used the definitions of the spherical functions in terms of the standard functions.

In order to study the defining equation (\ref{basic11}) on $\Lambda_m$, we define a
function $F_m^{(\nu)} (\lambda)$ on $\Lambda_0$ by
\beq\label{basic2}
F_m^{(\nu)}(\lambda) = \Sigma ( \lambda) J_{\nu }' ( \Sigma ( \lambda) ) H_{\nu }^{(1)} (e^{im \pi} \lambda )
 - e^{im \pi} \lambda J_{\nu} ( \Sigma  ( \lambda ))
H_{\nu}^{(1) '} ( e^{im \pi} \lambda  ) ,
\eeq
using the fact that $\Sigma ( e^{i m \pi} \lambda ) = \Sigma (\lambda)$, for $m \in \Z$.
It follows from
the fundamental equation (\ref{basic11})
that the zeros of $F_m^{(\nu)} (\lambda)$
on $\Lambda_0$ correspond to
the resonances of the one-dimensional Schr\"odinger operator
$H_\ell$ on the sheet $\Lambda_m$, for $|m| \geq 1$.
For $d$ odd, there are only two independent functions corresponding to $m=0$ and $m=-1$.
As $m=0$ corresponds to the physical sheet, there are no resonances, and
because $V_0 > 0$, there are no eigenvalues.
For $m= -1$, the resonances are the zeros of $F_1^{(\nu)}(\lambda)$ for $0 < \arg \lambda < \pi$
and $\nu $ a half-odd integer. We prove that this number is bounded below by $C_{-1} r^d$, with $C_{-1} > 0$.
For $d$ even, we will prove that the number of zeros is bounded below
by $C_m r^d$ on each nonphysical $m \neq 0$, for some constant $C_m > 0$.

The zeros of $F_m^{(\nu)}(\lambda)$, $\lambda \in \Lambda_0$, are studied using the uniform
asymptotic expansions of the Bessel and Hankel functions proved by Olver \cite{[Olver2],[Olver3]}.
A similar method was used by Stefanov \cite{stef06}. It is convenient
to define new variables $z = \lambda / \nu$ and $\tilde{z} (z) = (z^2 - \nu^{-2} V_0 )^{1/2}$.
The formulas for the analytic continuation of Bessel and Hankel functions \cite[chapter 7]{[Olver1]}
allow one to reduce the question of the zeros of $F_m^{(\nu)}(\lambda)$, $\lambda \in \Lambda_0$,
to considering those $\lambda \in \Lambda_0$ for which
\beq\label{eq:basic3}
{F}_0^{(\nu)} ( \nu z) = 2 m G_0^{(\nu)} ( \nu z) ,
\eeq
where, from (\ref{basic2}),
\beq\label{zero-sheet1}
F_0^{(\nu)} (\nu z) = \nu \tilde{z} J_\nu ' (\nu \tilde{z}) H_{\nu}^{(1)} (\nu z) -
 \nu z J_\nu (\nu \tilde{z}) H_{\nu}^{(1)'} (\nu z ) ,
\eeq
and we define
\beq\label{m-sheet21}
G_0^{(\nu)} ( \nu z) \equiv  \nu \tilde{z} J_\nu '
(\nu \tilde{z})  J_{\nu}  (\nu z )  - \nu z J_{\nu} (\nu \tilde{z} ) J_\nu^\prime (\nu z ) .
\eeq
It is sufficient for the lower bound to prove that
for any
$\nu < r$, for $\nu > \nu_0$ and
$r >> 0$ sufficiently large, that there are at least $\nu(1-\epsilon_1)$,
$\epsilon_1>0$ small, solutions
of the equation (\ref{eq:basic3}) in the half-disk
$\Im \lambda > 0$ and $| \lambda | \leq r$, uniformly in $r$ and $\nu$.

A special role in the uniform asymptotics of the Bessel and Hankel
functions is played by the compact, eye-shaped region $K$ in the complex
plane defined as follows.
Let $t_0$ be the positive root of $t = \coth t$, so $t_0 \sim 1.19967864 \ldots$. The region $K$ is
the symmetric region in the neighborhood of the origin bounded in $\C^+$ by the curve
\beq\label{k-region1}
z = \pm ( t \coth t - t^2)^{1/2} + i ( t^2 - t \tanh t)^{1/2}, ~0 \leq t \leq t_0,
\eeq
intercepting the real axis at $\pm 1$ and intercepting the imaginary
axis at $i z_0$, where $z_0 = (t_0^2 - 1)^{1/2} \sim 0.66274 \ldots$. The region $K$
is bounded by the conjugate curve in the lower half-plane.
To prove that there are at least $\nu(1-\epsilon_1)$ zeros
of $F_m^{(\nu)}(\lambda)$ near the upper boundary of the
eye-shaped region $\nu K$,
we concentrate a small region $\Omega_{1, \epsilon} \subset \C$, in the $z$ variable, near the upper
boundary of $K$, defined, for fixed $\epsilon>0$,
by $\Omega_{1, \epsilon}= \{ z\in \C^{+}:
\text{dist}\; (z,\partial K^+)<\epsilon\} \cap\{ z\in \C^{+}:\;
|z+ 1|>\epsilon\; \text{and}\; |z-1|>\epsilon\}$.

We computed the uniform asymptotics of each term in (\ref{eq:basic3}).
In order to state these, we need a map $z \rightarrow \rho(z)$ defined by
\beq\label{map1}
\rho (z) \equiv  \log \frac{1 + \sqrt{1 - z^2} }{z} - \sqrt{1 - z^2} .
\eeq
The uniform asymptotic expansion of $F_0^{(\nu)} (\nu z)$ for $z \in \Omega_{1, \epsilon}$,
as computed in \cite[sections 5 and 6]{ch-hi2}, is
\bea\label{mzeros01b}
F_0^{(\nu)} (\nu z) &=& \frac{- 2i}{\pi} \left\{ 1 - \frac{1}{\nu} \left[ \frac{V_0 (1-z^2)^{1/2}}{2 z^2}
\right]
+\mco\left(\frac{1}{\nu^2}\right) \right\},
\eea
and for $G_0^{(\nu)}$, we obtained
\beq\label{m-sheet3}
G_0^{(\nu)} (\nu z) =  \frac{e^{-2 \nu \rho}}{2 \pi} \left[ \frac{V_0 }{2 \nu^2 (1-z^2)} + \mathcal{O}
\left( \frac{1}{\nu^3} \right) \right] ,
\eeq
where the error is uniform for
$z\in \Omega_{1,\epsilon}$.
Consequently, the condition for zeros on the $m^{th}$-sheet is
that there exists solutions $z \in \Omega_{1, \epsilon}$ to the equation
\beq\label{reson-mth}
e^{2 \nu \rho (z) } \left( 1 + g_1(z, \nu)  \right)
= \frac{i m  V_0 }{4 \nu^2} \left( \frac{1}{1-z^2} \right) +
g_2(z, \nu),
\eeq
where $g_1(z,\nu)= \mathcal{O}(1/\nu)$, and $g_2(z,\nu)= \mathcal{O}(1/\nu^3)$, both uniformly for
$z\in \Omega_{1,\epsilon}$.
We note that for $V_0= 0$ there are no solutions to this equation.

We consider (\ref{reson-mth}) as an equation for $\rho$. The variable $\rho$ lies in a
set that is the image of $\Omega_{1, \epsilon}$ under the mapping
$z \rightarrow \rho$ given in (\ref{map1}). This set
contains a neighborhood of an interval of the negative
imaginary axis of the form
 $(-\pi+ h(\epsilon), -h(\epsilon))i
\subset (-\pi,0)i$. We proved that there exists at least $\nu (1- \epsilon_1)$ solutions in
a neighborhood of this set, where $\epsilon_1 = \mathcal{O} (\epsilon)$.
We first analyzed the zeros of the function $g(z,\nu)$ defined by
\beq\label{eq:gfnc1}
g(z , \nu ) = \nu^2 e^{2 \nu \rho} - \frac{i m V_0}{4}.
\eeq
These can be computed explicitly and have the form
\beq\label{g-zeros1}
\rho_k = \left\{ \frac{1}{2 \nu} \log \left( \frac{|m| V_0}{4} \right) - \frac{ \log \nu}{\nu} \right\}
+ i \frac{\pi}{\nu} \left[ k +
\operatorname{sgn}(m)\frac{1}{4} \right] , ~~ k \in \Z .
\eeq
Then, using Rouch\'e's Theorem, we proved that in a neighborhood of each
zero $\rho_k$ of $g(z, \nu)$ with imaginary part
in $( -i \pi + i2 h(\epsilon), -i2 h(\epsilon))$, there is exactly one solution
to (\ref{reson-mth}). Consequently,
there are at least $\nu(1-\epsilon_1)$, $\epsilon_1=\mco(\epsilon) >0$
zeros in a neighborhood of
the interval on the negative imaginary axis $( - \pi  , 0)i$ for all $\nu > 0$ large.

To prove the lower bound,
recall that $\nu=l+(d-2)/2$, $l \in \Natural$ and $\epsilon_1>0$ is arbitrary.
By the symmetry reduction described in the beginning of this section,
each zero of ${F}_m^{(\nu)}(\lambda)$ corresponds to a resonance of
multiplicity $m(l)$, the
dimension  of the space of spherical harmonics
on $\Sphere^{d-1}$ with eigenvalue $l(l+d-2)$.
Since $m(l)\geq cl^{d-2}+ \mathcal{O} (l^{d-3})$, for some $c>0$,
it follows that
\beq\label{count1}
n_{m,V}(r) \geq \sum_{\ell = 1}^{[r]}\frac{1}{4}(l-(d-2)/2)(cl^{d-2}+ \mathcal{O} (l^{d-3}))
\geq C_m r^{d}+ \mathcal{O} (r^{d-1}),
\eeq
for some $C_m >0$, depending on $m \in \Z^*$.
This proves the lower bound on the $m^{th}$-sheet, $m \in \Z^*$. If the potential $V$ is real,
the symmetry of the zeros means that
the resonances on $\Lambda_{-m}$ are in one-to-one correspondence with those on $\Lambda_m$.


\section{Open Problems}\label{sec:open}

One of the main open problems in this area is the proof of the optimal lower bound
$n_{V,m}(r) \geq C_{d,m} r^d$, $C_{d,m} > 0$,
for nontrivial, real-valued potentials $V \in L^\infty_0 (\R^d)$.
Alternately, it would be of interest to construct such a potential whose
resonance counting function has order
of growth strictly less than $d$ showing that such a lower bound is not possible.
The question of computing an asymptotic expansion for the resonance
counting function is of great interest but seemingly out of reach at this time.
It is not even clear if such an asymptotic
expansion should exist.


\section{Acknowledgements}
PDH thanks the organizing committee of 35th Journ\'ees EDP for the opportunity to present a
talk at Evian in June 2008.
TJC was partially supported by NSF grant 0500267, MSRI, and an MU Research Leave, and
PDH was partially supported by NSF grant 0503784,
during the time some of this work was done.


\end{document}